\documentclass[sigconf]{acmart}
\acmConference[EASE 2026]{The 30th International Conference on Evaluation and Assessment in Software Engineering}{9–12 June, 2026}{Glasgow, Scotland, United Kingdom}

\usepackage{multirow,makecell}
\usepackage{tcolorbox}
\usepackage{color,xcolor}
\usepackage{listings,amsfonts}
\usepackage{caption}
\usepackage{subcaption}
\usepackage{threeparttable}
\usepackage{bbding}
\usepackage{graphicx}
\usepackage{booktabs} 
\usepackage{longtable}
\usepackage{pifont}
\usepackage[linesnumbered,ruled,vlined]{algorithm2e}
\usepackage{tabularx}
\usepackage{enumerate}
\usepackage{enumitem}
\usepackage{amsmath}

\usepackage{amssymb}
\usepackage{soul, color, xcolor}
\usepackage{url}
\usepackage{hyperref}

\definecolor{customblue}{HTML}{006ca6}
\definecolor{customgreen}{HTML}{009264}
\definecolor{custombrown}{HTML}{ff3d00}
\AtEndPreamble{
 \usepackage{hyperref}
 \hypersetup{
  colorlinks = true,
  linkcolor = customblue,
  anchorcolor = customblue,
  citecolor = customgreen,
  filecolor = customblue,
  urlcolor = customblue
 }
}

\newcommand{\tool}[1]{\textsc{CommitSuite}}
\title{\tool{}: A Comprehensive Benchmark for Commit Classification and Message Generation}

\author[Z. Wan]{Zirui Wan}
\orcid{0009-0006-3782-7353}
\affiliation{%
  \institution{Huazhong University of Science and Technology}
  \city{Wuhan}           
  \country{China}
}
\email{ziruiwan@hust.edu.cn}

\author[Z. Wu]{Zhaonan Wu}
\orcid{0009-0002-9983-2407}
\affiliation{%
  \institution{Huazhong University of Science and Technology}
  \city{Wuhan}           
  \country{China}
}
\email{zhaonanwu@hust.edu.cn}

\author[X. Hou]{Xinyi Hou}
\orcid{0009-0005-9965-2109}
\affiliation{%
  \institution{Huazhong University of Science and Technology}
  \city{Wuhan}           
  \country{China}
}
\email{xinyihou@hust.edu.cn}

\author[Y. Zhao]{Yanjie Zhao}
\orcid{0000-0001-8793-5367}
\authornote{Corresponding author (Yanjie\_Zhao@hust.edu.cn).} 
\affiliation{%
  \institution{Huazhong University of Science and Technology}
  \city{Wuhan}           
  \country{China}
}
\email{Yanjie_Zhao@hust.edu.cn}

\author[P. Xia]{Pengcheng Xia}
\orcid{0009-0009-0201-3255}
\affiliation{%
  \institution{Huazhong University of Science and Technology}
  \city{Wuhan}           
  \country{China}
}
\email{xpc357@hust.edu.cn}

\author[H. Wang]{Haoyu Wang}
\orcid{0000-0003-1100-8633}
\affiliation{%
  \institution{Huazhong University of Science and Technology}
  \city{Wuhan}           
  \country{China}
}
\email{haoyuwang@hust.edu.cn}

\begin{document}

\begin{abstract}

High-quality commit messages are critical for maintaining software projects, yet ensuring their consistency and informativeness remains a practical challenge. While the Conventional Commits Specification (CCS) provides a structured format for commit messages, research on CCS-based commit classification and commit message generation (CMG) is limited by the absence of large-scale benchmarks, semantic annotations, and reliable evaluation methods. In this paper, we introduce \tool{}, a benchmark comprising 63,533 CCS-compliant commits from 243 open-source repositories across seven programming languages. Each commit is labeled with its CCS type and enriched with AST-level code changes, along with LLM-assisted semantic annotations that capture the ``what'' and ``why'' behind the change. To evaluate CMG systems, we propose a reference-free framework based on five binary metrics: rationality, comprehensiveness, non-redundancy, authenticity, and logicality, enabling semantic-level assessment without relying on human-written references. Our experiments show that LLMs can effectively support both generation and evaluation, with evaluation achieving 0.849 Cohen's Kappa agreement against human judgments. \tool{} offers a unified resource for structured commit understanding and facilitates reproducible research on commit classification and generation.

\end{abstract}

\maketitle

\section{Introduction}
\label{sec:introduction}

Commit messages play a fundamental role in software development workflows. In version control systems such as Git, a \textit{commit} records an atomic change to the codebase, including both the code modifications and a message that conveys the intent and rationale behind the change. Besides, commits form the historical record of a project, allowing developers to trace code evolution, understand past decisions, and collaborate effectively in distributed environments~\cite{li2023commit}.
Well-written commit messages enhance code maintainability, facilitate debugging and impact analysis, and improve team communication~\cite{santos2016judging}. They provide essential context for future readers, aid in fault localization, and document the development process in a reusable form. In large-scale or long-lived projects, commit messages serve as a critical knowledge base that complements the source code. As such, \textbf{commit message generation (CMG)} has become an active research area focused on automatically producing high-quality messages that reflect both the content and the intent of code changes.
In practice, commit messages are often inconsistent, lack detail, or are omitted due to time pressure or unclear guidelines~\cite{li2023commit}. This reduces the usefulness of commit histories and hinders tasks such as code review, changelog generation, and semantic versioning - a standardized approach where commit types (e.g., \texttt{feat}, \texttt{fix}) automatically determine version number increments (major.minor.patch)~\cite{buse2010automatically,de2009software,conventionalcommitsConventionalCommits}. 

Aiming to improve the structure and quality of commit messages, the \textbf{Conventional Commits Specification (CCS)}~\cite{conventionalcommitsConventionalCommits} has been widely adopted in software development. CCS enforces a structured format, typically \texttt{<type>(<scope>): <description>}, where \texttt{type} (e.g., \texttt{feat}, \texttt{fix}) indicates the nature of the change and \texttt{scope} specifies the affected component. This format enhances both human readability and machine interpretability, enabling automated tools for commit classification, changelog generation, and semantic versioning. Currently, numerous high-impact repositories adopt CCS, including \href{https://github.com/puppeteer/puppeteer}{\texttt{puppeteer/puppeteer}} (91.6k stars), \href{https://github.com/shadcn-ui/ui}{\texttt{shadcn-ui/ui}} (92.1k stars), and \href{https://github.com/gatsbyjs/gatsby}{\texttt{gatsbyjs/gatsby}} (55.9k stars). This widespread adoption underscores CCS's vital role in establishing standardized, machine-interpretable commit practices essential for modern software collaboration. 

Although CCS provides a clear structure for commit messages, research on CCS-based commit classification and CMG still faces several key challenges.
One major limitation is the \textbf{lack of large-scale, high-quality benchmarks and tools} for fine-grained classification across the full CCS type set~\cite{sarwar2020multi, zeng2024first}. Most existing work focuses on binary or coarse-grained categories, leaving the more nuanced classification tasks underexplored.
In addition, many CMG datasets suffer from \textbf{low standardization and lack semantic annotations} that capture both the ``what'' and ``why'' of code changes. They also often omit fine-grained contextual information, such as function-level or structural representations, which are essential for generating accurate and informative messages.
Finally, \textbf{evaluation remains a major bottleneck}. Current methods rely heavily on reference-based metrics like BLEU and ROUGE-L, which are highly sensitive to inconsistencies in human-written messages. These metrics often fail to reflect semantic correctness, informativeness, or adherence to CCS format~\cite{tsvetkov2024towards, li2024only}, limiting the reliability and comparability of CMG systems.

To address these limitations, we present \tool{}, a large-scale benchmark for fine-grained CCS-based commit classification and message generation. It comprises 63,533 high-quality commits from 243 open-source repositories, all strictly following the CCS format and covering seven popular programming languages. Each commit is labeled with its CCS type and enriched with function-level and AST-level structural changes. We also use large language models (LLMs) to semi-automatically annotate each commit with ``what'' and ``why'' semantics to support more informative generation. Beyond providing this comprehensive dataset and structural annotations, we propose a reference-free evaluation framework with binary metrics, which consist of \textit{rationality}, \textit{comprehensiveness}, \textit{non-redundancy}, \textit{authenticity} and \textit{logicality}. We show that LLMs can reliably perform these evaluations, enabling scalable and objective assessment of CMG systems.
We further utilize \tool{} to conduct extensive evaluations, benchmarking the performance of state-of-the-art commit message generation tools and leading LLMs on both commit classification and message generation tasks. Our main contributions are as follows:

\begin{itemize}[leftmargin=1em]
  \item \textbf{A novel, high-quality benchmark dataset.}  
  We introduce \tool{}, comprising 63,533 strictly CCS-compliant commits (243 repos, 7 languages) enriched with AST-level changes and LLM-assisted ``what''/``why'' annotations.

  \item \textbf{A reference-free CMG evaluation framework.}  
  We propose five binary metrics (rationality, comprehensiveness, non-redundancy, authenticity, logicality) for semantic assessment without human references, validated via LLM automation.

  \item \textbf{Comprehensive performance benchmarking. }
Evaluations using \tool{} show LLMs significantly outperform traditional tools in CMG (especially on semantic metrics), while highlighting challenges in fine-grained classification. \tool{}'s high quality and CCS alignment enable reliable and reproducible benchmarking. 

  \item \textbf{Open-source and reproducibility.}  
  All datasets, evaluation tools, and baseline results are publicly released to support reproducible research at: \url{https://github.com/security-pride/CommitSuite}.
\end{itemize}

\section{Background and Motivation}
\label{sec:background}

\subsection{Commit Message}
\label{sec:cmg_background}

\subsubsection{Commit Message Generation (CMG)}
A \textit{commit} represents a recorded change to one or more files in a code repository along with a descriptive message, known as \textit{commit message}. The actual modifications are documented in the \textit{diff} section, which highlights added, removed, or altered lines of code. CMG refers to the automated generation of commit messages from code diffs and has become an important technique in modern software development. High-quality commit messages are essential for helping developers understand both the content and motivation behind code changes, thereby facilitating effective collaboration and long-term maintenance.
Traditional CMG methods, such as retrieval-based and learning-based approaches, often struggle to produce messages that are accurate, complete, and readable~\cite{wang2021quality}. With the rapid development of large language models (LLMs), there is growing interest in applying LLMs to code-related tasks, including CMG. Recent studies~\cite{xue2024automated, fan2024exploring, zhang2024using, li2024only, zhang2024rag} have shown that LLMs excel at generating informative and context-aware commit messages, due to their advanced language understanding and generation capabilities.

\subsubsection{Conventional Commits Specification (CCS)}

\begin{figure}[ht]
    \centering    \includegraphics[width=\linewidth]{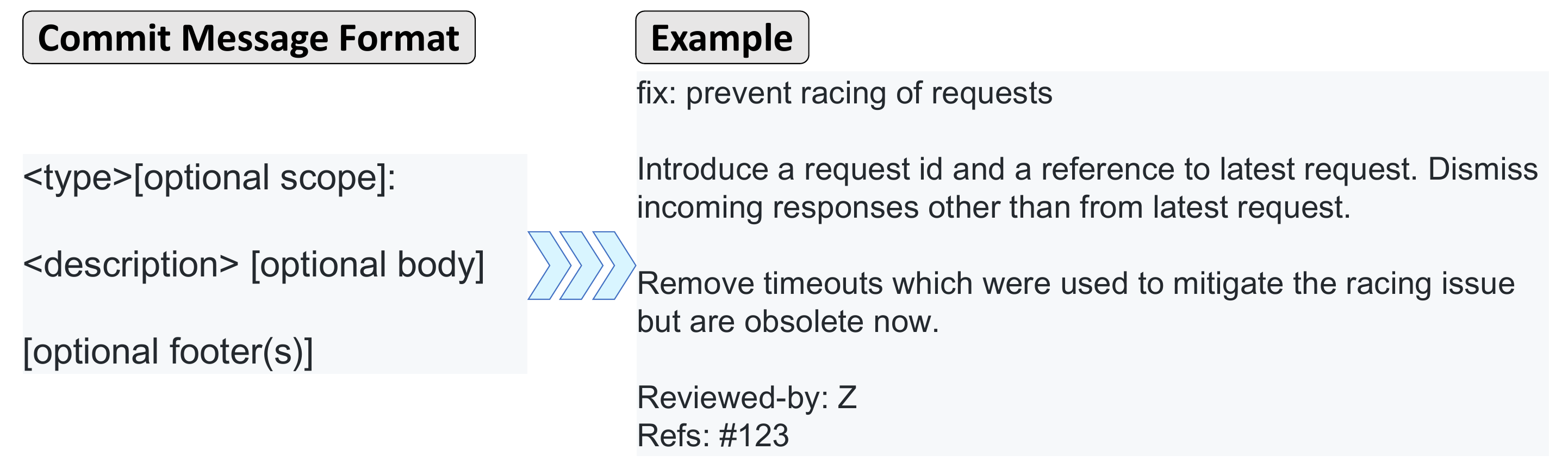}
    \caption{Commit message format defined by CCS.}
    \label{fig:msg_example}
\end{figure}

CCS v1.0.0 was released in 2019~\cite{conventionalcommitsConventionalCommits, githubGitHubConventionalcommitsconventionalcommitsorg}, which provides a standardized convention for writing commit messages, aiming to make them easily understood by both humans and machines. CCS defines a specific message format, as shown in \autoref{fig:msg_example}, which requires each commit message to include fields such as \verb|<type>|, \verb|<scope>|, and \verb|<description>|. Among these, the \verb|<type>| field describes the nature of the code change, with common values including \verb|feat| (new feature), \verb|fix| (bug fix), and \verb|docs| (documentation update), among others.The primary goal of CCS is to enhance the readability and maintainability of code changes by enforcing a consistent structure for commit messages. In recent years, CCS has gained widespread adoption in the open-source community and has become a widely recognized standard for commit messages. Building on this, Zeng et al.~\cite{zeng2024first} refined and standardized commit classification criteria, as summarized in \autoref{tab:commit-types}. The ten-category classification task can be viewed as a subtask of CMG, highlighting the need to integrate CCS specification checks into the construction of mainstream CMG datasets. 

\begin{table}[h!]
\centering
\caption{Ten-category standards proposed by Zeng et al.~\cite{zeng2024first}}
\resizebox{1\linewidth}{!}{
\begin{tabular}{cl}
\toprule[1.2pt]
\textbf{Type}   & \textbf{Description} \\
\midrule[1.2pt]
feat      & Introduce new features to the codebase. \\
fix       & Fix bugs or faults in the codebase. \\
perf      & Improve performance, such as speed or memory usage. \\
style     & Improve code readability without changing functionality. \\
refactor  & Restructure code without changing its behavior. \\
docs      & Modify documentation or comments. \\
test      & Add or update tests. \\
ci        & Change CI configuration or scripts (e.g., CI/CD pipelines). \\
build     & Modify the build system or dependencies. \\
chore     & Miscellaneous changes not fitting other categories. \\
\bottomrule[1.2pt]
\end{tabular}}
\label{tab:commit-types}
\end{table}

\begin{table*}[t!]
  \centering
  \caption{Comparison of different commit message datasets.}
  \small
  \renewcommand{\arraystretch}{1}
  \setlength{\tabcolsep}{3pt}
  \resizebox{0.8\linewidth}{!}{
  \begin{tabular}{@{}ccccccccc@{}}
    \toprule[1.2pt]
    \textbf{Feature} & \textbf{CommitChronicle} & \textbf{CommitBench} & \textbf{MCMD} & \textbf{FIRA} & \textbf{CommitBERT} & \textbf{OMG} & \textbf{CMO} & \textbf{CommitSuite (ours)} \\
    \midrule[1.2pt]
    Size & 10,700,000 & 1,664,590 & 2,250,000 & 90,661 & 345,759 & 35,431 & 500 & 63,533 \\
    
    Repositories & 11,900 & 72,000 & 500 & 1,000 & 52,462 & 32 & 32 & 172 \\
    
    Languages & \makecell{20\\languages} & \makecell{Java, Ruby,\\JavaScript, Go,\\PHP, Python} & \makecell{C++, C\#,\\Java, Python,\\JavaScript} & Java & \makecell{Java, Ruby,\\JavaScript, Go,\\PHP, Python} & Java & Java & \makecell{C, C++, Java,\\Python, Go,\\JavaScript, TypeScript} \\[11pt]
    
    License Filtering & \makecell{Apache-2.0,\\MIT, BSD-3-Clause} & MIT & -- & -- & \makecell{Re-distribution\\license} & \makecell{Apache\\projects} & \makecell{Apache\\projects} & \makecell{Apache-2.0,\\MIT, BSD-3-Clause} \\
    
    File-type Check & × & \checkmark & \checkmark & × & \checkmark & \checkmark & \checkmark & \checkmark \\
    
    Bot Check & \checkmark & \checkmark & \checkmark & × & × & \checkmark & \checkmark & \checkmark \\
    
    Duplicate Check & \checkmark & \checkmark & \checkmark & \checkmark & × & \checkmark & \checkmark & \checkmark \\
    
    Issue/PR & × & × & × & × & × & \checkmark & \checkmark & \checkmark \\
    
    What/Why Information & × & × & × & × & × & × & \checkmark & \checkmark \\
    
    AST Change Analysis & × & × & × & \checkmark & × & × & \checkmark & \checkmark \\
    
    CCS Check & × & × & × & × & × & × & × & \checkmark \\
    \bottomrule[1.2pt]
  \end{tabular}}
  \label{tab:transposed_datasets}
\end{table*}

\subsection{Commit Datasets and Their Limitations}

\autoref{tab:transposed_datasets} compares existing datasets with \tool{}. CommitChronicle~\cite{eliseeva2023commit} provides scale (10.7M commits, 20 languages) but lacks quality annotations and CCS compliance. CommitBench~\cite{schall2024commitbench} filters 1.7M messages by license and bot detection yet omits issue links and quality evaluation. MCMD~\cite{tao2021evaluation} covers 2.25M messages from top-starred projects but excludes AST analysis. FIRA~\cite{schall2024commitbench} enriches 90K Java commits with fine-grained AST edits yet is limited to a single language. CommitBERT~\cite{jung2021commitbert} selects 345K simple commits with partial CCS filtering but limited context. OMG~\cite{li2024only} collects 35K Java commits with issue/PR links and manual/automatic evaluation, yet lacks ``what/why'' annotations. CMO~\cite{li2025consider} provides 500 Java commits with issue/PR links, ``what/why'' labels, and manual quality ratings, but its small scale limits generalizability. None combine strict CCS compliance, multi-language coverage, semantic annotations, and AST-level context—gaps \tool{} addresses.

 \begin{figure}[ht]
    \centering    \includegraphics[width=\linewidth]{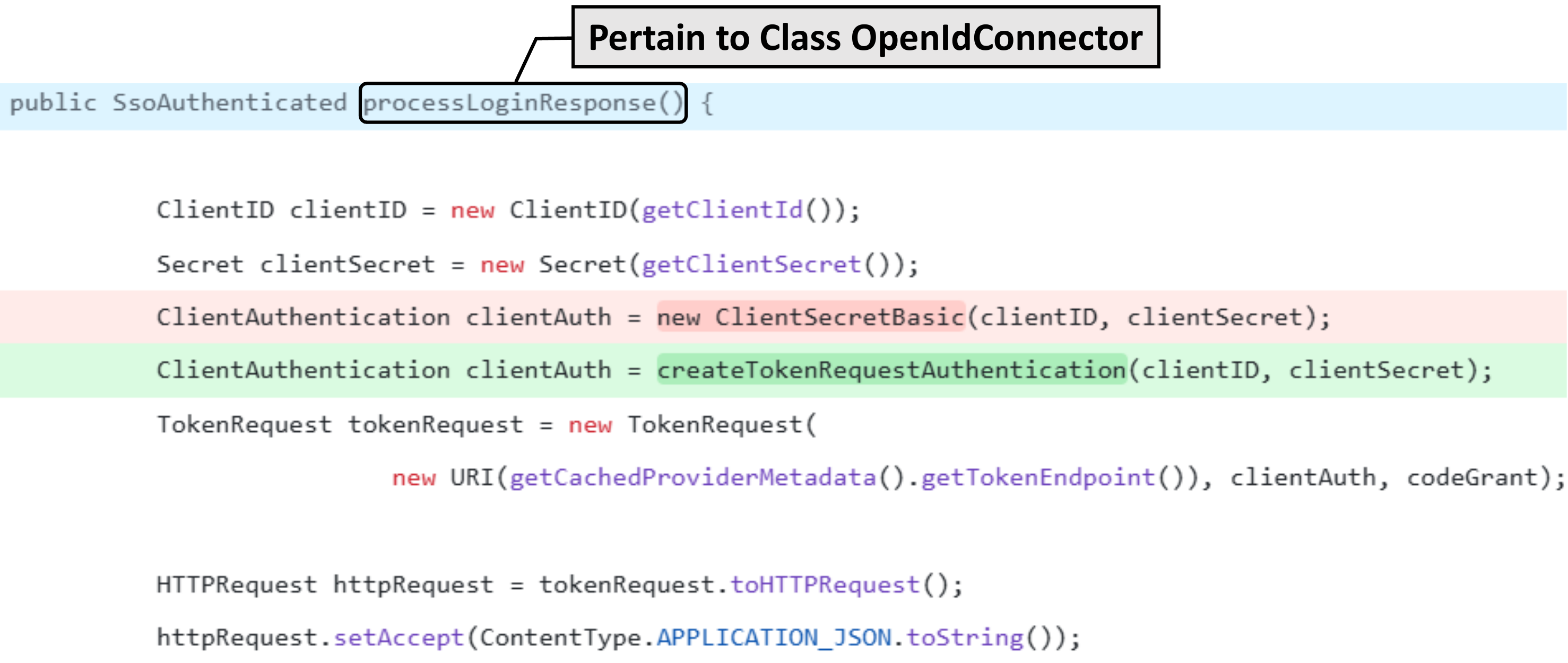}
    \caption{Missing function-level information in diff.}
    \label{fig:connect}
\end{figure}

\begin{figure*}[t!]
    \centering    \includegraphics[width=0.8\linewidth]{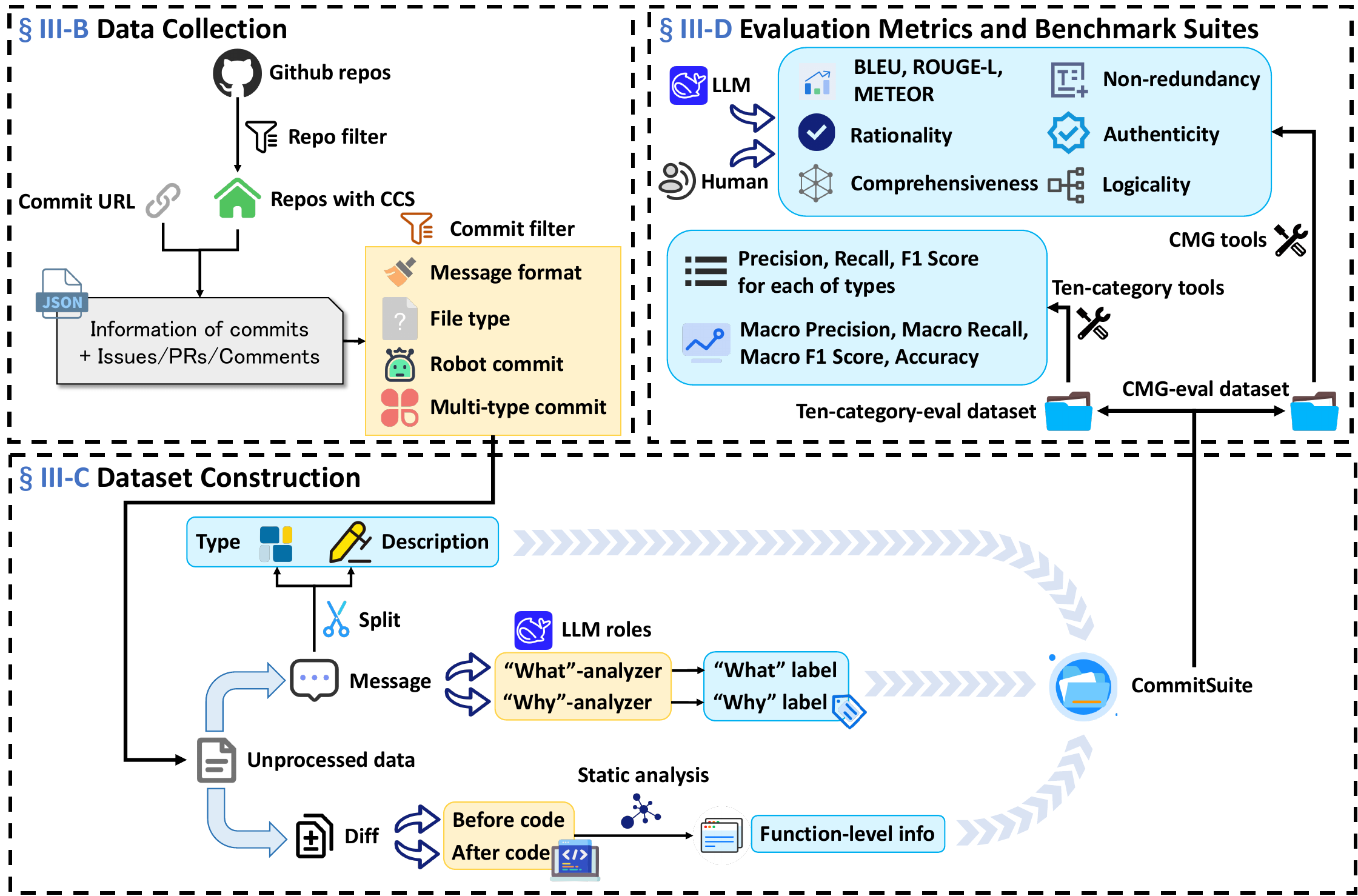}
    \caption{Overview of \tool{}.}
    \label{fig:overview}
\end{figure*}

\subsection{Challenges in CMG Evaluation and Data Quality}

\textbf{Limitations of Reference-Based Metrics.}
Most current CMG tools rely on traditional machine translation metrics such as BLEU, ROUGE-L, and METEOR for evaluation. However, since developer-written reference messages lack strict standards and vary significantly in quality, these metrics are often unreliable for assessing generation performance~\cite{li2024only}. To address this issue, a high-quality evaluation dataset or a reference-free evaluation standard is required. 

\textbf{Limitations of Previous Dataset.} While previous datasets have advanced the field, they remain insufficient for our focus on fine-grained CCS-based commit classification and semantically rich message generation. Key limitations include:

1) widespread lack of strict CCS compliance, preventing reliable benchmarking of classification tasks and hindering research on semantic versioning workflows; 

2) absence of systematic ``what''/``why'' annotations, which not only compromises reference-based evaluation but also limits the availability of high-quality training data; 

3) most datasets focus solely on commit diffs, which omit critical contextual information such as enclosing class structures and data dependencies. For example, as illustrated in \autoref{fig:connect}, the diff may indicate the modified function but omit critical function-level context such as its enclosing class structure and data dependencies, leading to information loss and reduced semantic understanding by LLMs.

\newcommand{\Dall}{\mathcal{D}_{\text{all}}}
\newcommand{\Dhuman}{\mathcal{D}_{\text{human}}}
\newcommand{\Dgen}{\mathcal{D}_{\text{gen}}}
\newcommand{\Dcls}{\mathcal{D}_{\text{cls}}}
\newcommand{\Dastcmg}{\mathcal{D}_{\text{ast}}^{\text{cmg}}}
\newcommand{\Dastten}{\mathcal{D}_{\text{ast}}^{\text{ten}}}
\newcommand{\Dcmg}{\mathcal{D}_{\text{cmg}}}
\newcommand{\Dten}{\mathcal{D}_{\text{ten}}}

\section{Approach}
\label{sec:approach}

\subsection{Overview}

\autoref{fig:overview} illustrates the overall architecture of \tool{}. It includes three parts: \textit{data collection} (\S\ref{sec:data_collection}), \textit{dataset construction} (\S\ref{sec:data_processing}), and \textit{evaluation metrics and benchmark suites} (\S\ref{sec:evaluation_module}).

\textbf{In the data collection phase}, we identified GitHub repositories adhering to CCS via their mentions of the CCS website, then crawled commits from them. At last, we filtered these commits by various rules, resulting in 63,533 commits. \textbf{In the dataset construction phase}, commit messages were systematically divided into two components: the \textit{type} and the \textit{description}. Then we manually verified the used \textit{what} and \textit{why} information annotated by a LLM. 
Additionally, we used a static analysis tool to integrate function-level change information into the dataset. \textbf{For evaluation metrics and benchmark suites}, we selected test sets from \tool{} for the commit classification task and the CMG task. Both tasks were evaluated using traditional metrics as well as newly proposed metrics.

\begin{figure*}[t!]
    \centering    \includegraphics[width=\linewidth]{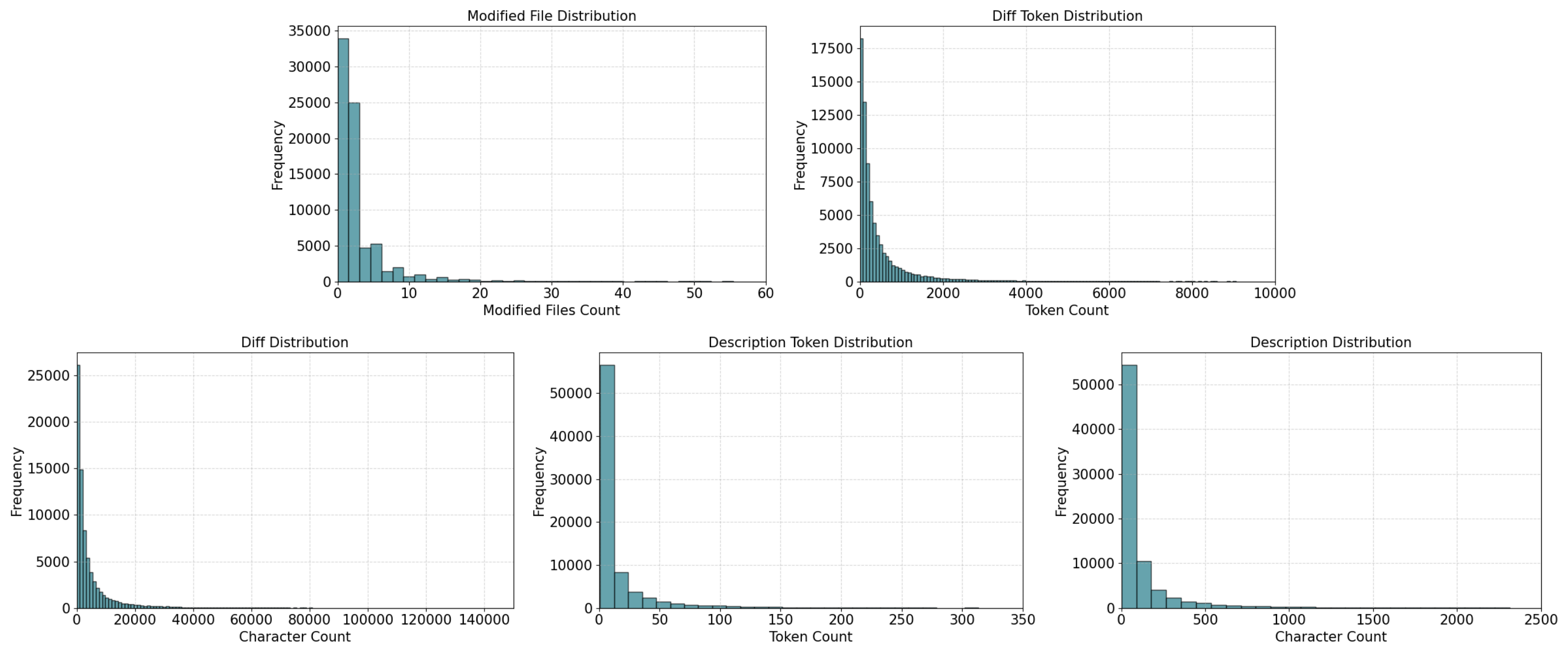}
    \caption{The distributions of ``Diff length'', ``Description character count'', ``Diff token count'', ``Description token count'', and ``Number of modified files''.}
    \label{fig:distribution}
\end{figure*}

\subsection{Data Collection}
\label{sec:data_collection}

\subsubsection{\textbf{Repository Selection}}
\label{sec: rep_select}

Analysis of top GitHub repositories (ranked by stars) reveals that projects maintaining CCS-compliant commits consistently mention the official CCS website  (\href{https://conventionalcommits.org/}{conventionalcommits.org}) in their documentation files like \texttt{README.md} or \texttt{CONTRIBUTING.md}. According to this feature, we identified 243 repositories covering seven common languages: C, C++, Java, Python, Go, JavaScript, and TypeScript. All these repositories were created between 2 and 10 years ago, which use permissive licenses (Apache-2.0, MIT, BSD-3-Clause), have $\geq$10 forks (indicating high community engagement) and have an average daily commit count $<$10 after 2020 (to avoid massive robot commits or low-quality commits; since CCS v1.0.0 was released in 2019, we only collect data from 2020 onwards). Their median star count is $1,669$, indicating high recognition and popularity in the developer community.
Based on the initial search, we performed manual screening. First, we excluded 54 repositories whose commits almost entirely violated the CCS specification (i.e., messages did not follow CCS format or used non-CCS types). Next, we flagged 49 repositories that were mostly CCS-compliant but contained a small fraction of non-conforming commits; all commits from these repositories were retained and explicitly annotated as ``partially compliant” during downstream processing. The remaining 140 repositories were confirmed to be strictly CCS-compliant.

\subsubsection{\textbf{Data Crawling}}

In this step, our goal is to collect up-to-date and comprehensive commits. We collected commits up to May 5, 2025 from the selected repositories  using PyDriller~\cite{spadini2018pydriller} and GitHub API~\cite{githubGitHubREST}, recording each commit's hash, message, author, email, date, modifications, associated comments and related issues/PRs. After above steps, a total of 358,921 commits were obtained.

\subsubsection{\textbf{Data Filtering}}
To ensure the quality of the dataset, we adopted the following rules to filter commits:
\begin{itemize}
    \item We used regular expressions to filter commit messages conforming to the CCS format, removing 124,829 commits (34.78\% of the total).
    \item We filtered by file extensions to retain only the seven languages mentioned earlier; commits touching files of multiple languages were discarded. This step removed 152,596 commits (65.19\% of the remaining data).
    \item To avoid spurious patterns,  we removed commits with authors or emails containing ``bot'', ``robert'', or ``b0t'' (case-insensitive) and further filtered using known robot lists~\cite{chidambaram2023dataset,golzadeh2021ground}, removing 1,141 commits (1.40\% of the remaining data).
    \item A commit was labeled “multi-type” if its message contained two or more CCS type keywords (e.g., ``fix:... feat: …”); such commits were excluded. This step removed 2,458 commits (3.06\% of the remaining data).

\end{itemize}

Finally, 77,897 unique commits were retained. After inspection, there are no duplicates in the data.

\subsubsection{\textbf{Outlier Removal}}
To exclude atypical or memory-intensive data, we analyzed distributions of ``Diff length'', ``Description character count'', ``Diff token count'', ``Description token count'', and ``Number of modified files''. As shown in \autoref{fig:distribution},  these metrics did not follow normal distributions, so we used the IQR (Interquartile Range) method~\cite{tukey1977exploratory} to remove extreme outliers, eliminating 14,364 commits (18.44\% of the remaining data). The final dataset has 63,533 commits from 172 repositories, covering 7 languages.

\subsection{Dataset Construction}
\label{sec:data_processing}

\begin{figure}[ht]
    \centering    \includegraphics[width=\linewidth]{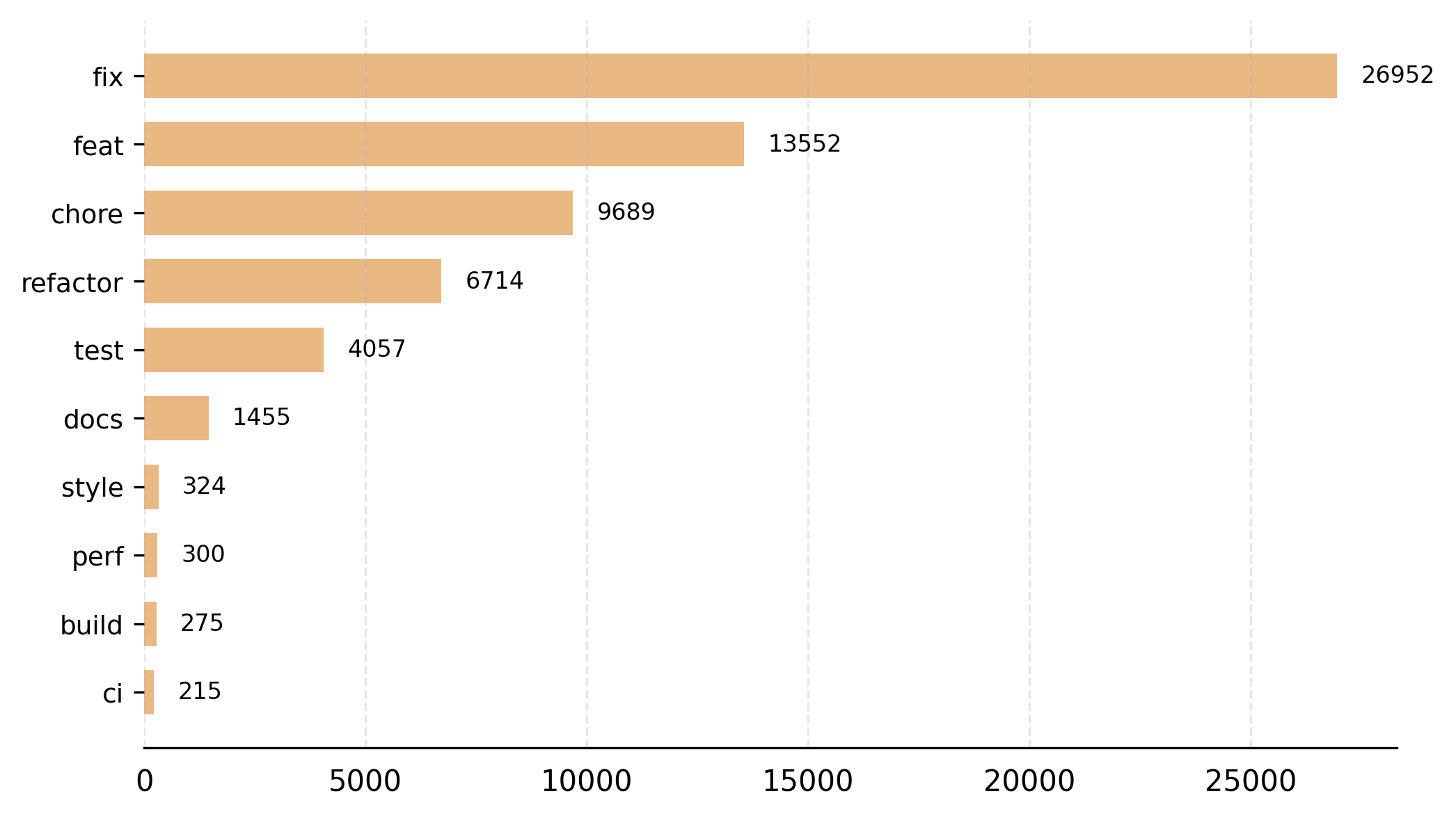}
    \caption{Commit type distribution in \tool{}.}
    \label{fig:type_distribution}
\end{figure}

\begin{figure}[ht]
    \centering    \includegraphics[width=0.8\linewidth]{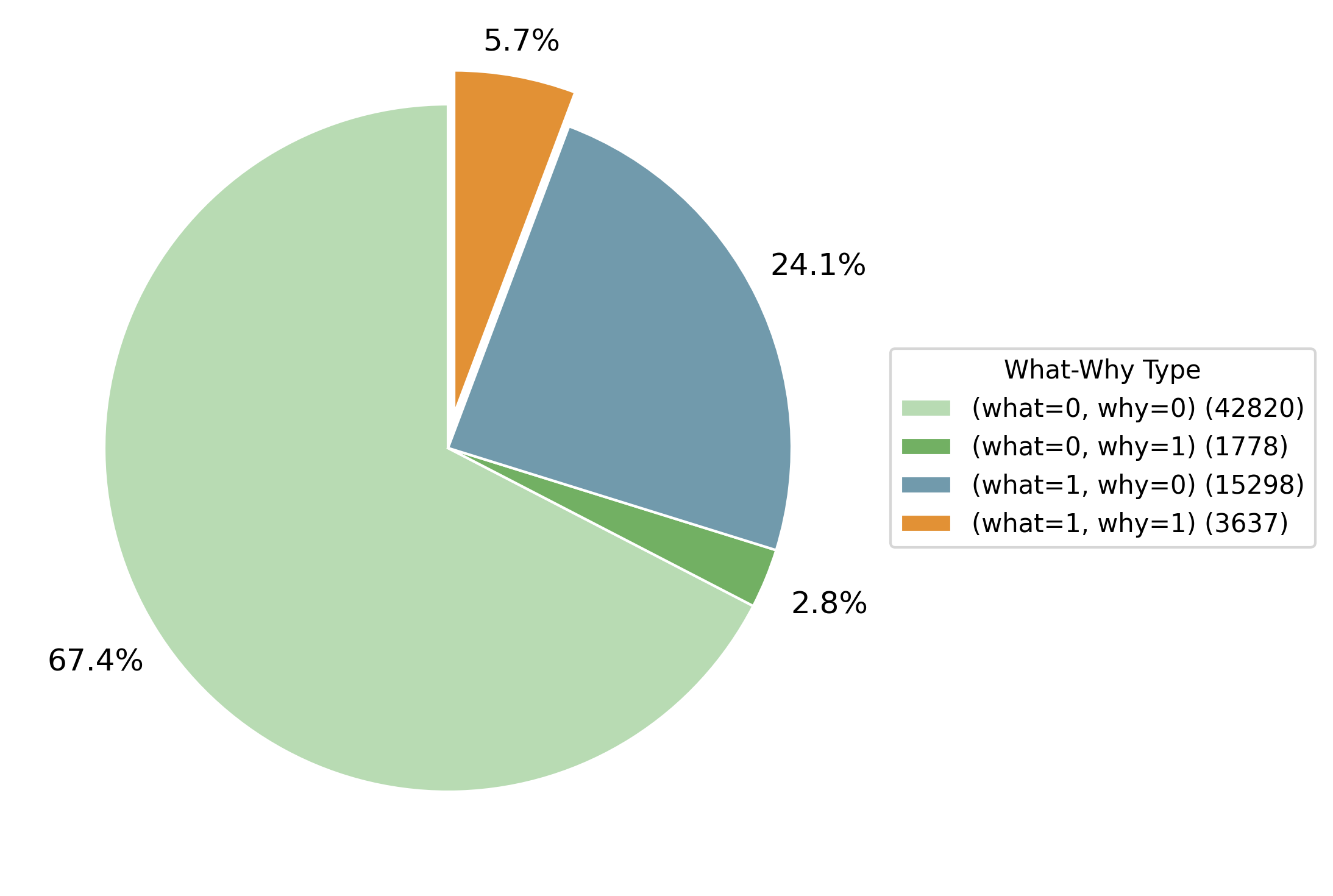}
    \caption{The distribution of commit messages categorized by the presence of ``what'' and ``why'' information in \tool{}.}
    \label{fig:what_why_distribution}
\end{figure}

Firstly, we parse each commit message to extract the type prefix and description text (discarding any scope information) using regular expressions. The distribution of the ten classification types is shown in \autoref{fig:type_distribution}.
After that, we assess the entire message content for the presence of ``what'' and ``why'' information as defined by Tian et al.~\cite{tian2022makes}. The message quality classifiers proposed in their research have low precision in ``good messages'' (include both ``what'' and ``why'') and Xue et al.~\cite{xue2024automated} also demonstrated in their research that LLMs exhibit high consistency with humans in this classification task. Therefore, we used the DeepSeek-V3 model to annotate message quality for its cost-effective. With distributions shown in \autoref{fig:what_why_distribution}, data was divided into four annotation states based on presence flags:

\begin{itemize}
    \item 00: Both ``what'' and ``why'' absent
    \item 01: ``Why'' present, ``what'' absent
    \item 10: ``What'' present, ``why'' absent
    \item 11: Both present
\end{itemize}
 
Three researchers manually verified 182 ``good messages'' (5\% of messages with both ``what'' and ``why''), achieving 93.41\% precision. Additionally, on 500 randomly sampled commits, we measured accuracies of 0.854 (what) and 0.966 (why) with an average Cohen's Kappa of 0.834. Despite this quality annotation, all commits—including those lacking ``what” or ``why” information—are retained in the dataset to ensure its completeness for the ten-category classification task.

To provide richer CMG context, we used tree-sitter~\cite{treesitterIntroductionTreesitter,githubGitHubConventionalcommitsconventionalcommitsorg} to compute AST-level diffs between pre- and post-commit versions, capturing added, modified, or deleted functions, classes, and other structures (\autoref{tab:ds-changes}). By aligning tree-sitter node spans with diff line numbers, we linked every code change to its enclosing function or class, yielding concise, function-level context for each commit. 

\begin{table}[htbp]
\centering
\caption{Detected data structure changes by programming language.}
\renewcommand{\arraystretch}{1.1}
\label{tab:ds-changes}
\resizebox{0.95\linewidth}{!}{
\begin{tabular}{ll}
\toprule[1.2pt]
\textbf{Language} & \textbf{Detected Data Structures} \\
\midrule[1.2pt]
C/C++ & Function, Class, Struct, Enum, Namespace \\
Java & Method, Class, Interface, Enum, Annotation \\
Python & Function, Class, Async Function \\
Go & Function, Struct, Interface \\
JavaScript & Function, Class, Method, Arrow Function, Object \\
TypeScript & \makecell[l]{Function, Class, Method, Interface,\\Type Alias, Enum, Arrow Function} \\
\bottomrule[1.2pt]
\end{tabular}}
\end{table}

\subsection{Evaluation Metrics and Benchmark Suites}
\label{sec:evaluation_module}

To demonstrate \tool{}’s utility as a benchmark, we further evaluate both (1) automatically generated commit messages from state-of-the-art CMG tools/LLMs and (2) CCS-based ten-category classification performance, using the evaluation suites defined below.

For CCS-compliant messages, the ``type'' and ``description'' (content after the colon) should be checked separately. To ensure balanced representation across different languages and commit types for fair evaluation, we constructed two specialized evaluation subsets from \tool{}: 

\begin{enumerate}
    \item \textbf{Ten-category-eval Dataset} ($\Dten$): Due to the limited availability of certain types—particularly \textit{ci}, which had only 116 valid samples after filtering—we selected 116 correctly classified samples for each category (1,160 in total). While preserving developers' original classification labels, we rigorously excluded commits where the CCS type did not accurately reflect the code changes. This ensures high label reliability while maintaining authentic classification practices. 

    \item \textbf{CMG-eval Dataset} ($\Dcmg$): We randomly sampled 1,000 commits whose messages include  ``what'' and ``why'' from \tool{} (180 for C/C++, 102 for Java, 180 for Python, 180 for Go, 179 for JavaScript, 179 for TypeScript).

\end{enumerate}

These evaluation subsets are designed specifically for controlled benchmarking tasks. As the largest and only fully CCS-compliant commit dataset currently available (63,533 commits with 100\% format verification), the full \tool{} provides unique value for broader research and tool development, enabling studies on real-world CCS adoption patterns, training of CCS-aware versioning and changelog systems, and analysis of structured commit practices across diverse software ecosystems.

Since \tool{} already filters out commits that modify files in multiple programming languages during the data collection phase (see §\ref{sec:data_collection}), all commits in the evaluation datasets modify files exclusively in a single language. This allows evaluation to be performed on a per-language basis as needed.
For these two tasks, our evaluation criteria are as follows:

    1) \textbf{Classification Evaluation}: 

Similar to other multi-classification tasks~\cite{ghamrawi2005collective, zhang2013review}, classification performance is measured by precision, recall, and F1-score.

    2) \textbf{Message Quality Evaluation}: 
    
    Traditional text generation tasks often use three metrics~\cite{schall2024commitbench}: Traditional metrics include BLEU (n-gram similarity with brevity penalty)~\cite{papineni2002bleu}, ROUGE-L (LCS-based content overlap)~\cite{lin2004rouge}, and METEOR (stem/synonym matching with word order penalties)~\cite{banerjee2005meteor}.

These metrics measure similarity to references, but in CMG tasks, reference quality is highly variable due to lax commit message standards. References often contain non-standard content (e.g., URLs, non-English text), compromising metric reliability. This necessitates reference-free evaluation standards.

Li et al.~\cite{li2024only} pioneered reference-free commit message evaluation using manual metrics scored via a 5-point Likert scale ~\cite{likert1932technique}. However, their follow-up study ~\cite{li2025consider} identified two key limitations: 1) Limited automation capability (LLM Macro-F1=0.766); 2) Intrinsic metric conflicts where optimizing one metric degrades another. To overcome these constraints, we propose an enhanced binary metric system defined in \autoref{tab:metrics}.

\begin{table}[ht]
\centering
\caption{\tool{}'s commit message quality metrics.}
\label{tab:metrics}
\renewcommand{\arraystretch}{1.1}
\begin{tabularx}{\linewidth}{>{\bfseries}l X}
\toprule[1.2pt]
\textnormal{\textbf{Metric}} & \textbf{Description} \\
\midrule[1.2pt]
Rationality & Whether it contains ``why” information. \\
\midrule
Comprehensiveness & Whether it contains ``what” information and covers all affected files. \\
\midrule
Non-redundancy & Whether there is no semantic repetition, mergeable details, meaningless content (unrelated to ``what” and ``why”), or content of little use.\\
\midrule
Authenticity & Whether the content in it does not include modifications absent in the actual code changes.\\
\midrule
Logicality & Whether the content in it is reasonable and logical.
\\
\bottomrule[1.2pt]
\end{tabularx}
\end{table}

Our new metrics use binary evaluations. Compared to the evaluation criteria proposed by Li et al.~\cite{li2024only}, this evaluation method is more suitable for LLM-powered automated assessment. We only need to calculate the proportion of responses to evaluate tool performance. We replaced Conciseness with Non-redundancy to avoid the trade-off between comprehensiveness and conciseness for complex commits. The new Authenticity and Logicality metrics address LLM hallucinations, preventing fictional or illogical content. As for the original Expressiveness indicator, it is indeed difficult to judge it with binary evaluation. However, in Li et al.'s experimental results~\cite{li2025consider}, the consistency between LLMs' scoring results for this indicator and manual scoring results was relatively high (F1 score of 0.875). Therefore, if evaluation of this indicator is required, we can retain the 5-point Likert scale scoring method and use LLMs or manual methods for scoring. 

\providecommand{\Dall}{\mathcal{D}_{\text{all}}}
\providecommand{\Dhuman}{\mathcal{D}_{\text{human}}}
\providecommand{\Dgen}{\mathcal{D}_{\text{gen}}}
\providecommand{\Dcls}{\mathcal{D}_{\text{cls}}}
\providecommand{\Dastcmg}{\mathcal{D}_{\text{ast}}^{\text{cmg}}}
\providecommand{\Dastten}{\mathcal{D}_{\text{ast}}^{\text{ten}}}
\providecommand{\Dcmg}{\mathcal{D}_{\text{cmg}}}
\providecommand{\Dten}{\mathcal{D}_{\text{ten}}}

\begin{table}[ht]
\centering
\caption{Dataset subsets used in each RQ.}
\label{tab:rq_datasets}
\renewcommand{\arraystretch}{1.15}
\small
\begin{tabular}{@{}p{0.15\linewidth}p{0.2\linewidth}p{0.15\linewidth}p{0.4\linewidth}@{}}
\toprule
\textbf{RQ} & \textbf{Subset} & \textbf{Size} & \textbf{Derived from} \\
\midrule
RQ1 & $\Dhuman$ & $300$ & $\Dhuman\subset\Dcmg$  \\
RQ2 & $\Dcmg$   & $1000$ & $\Dcmg\subset\Dall$  \\
RQ3 & $\Dten$   & $1160$ & $\Dten\subset\Dall$  \\
RQ4 & $\Dastcmg$, $\Dastten$    & $200$, $200$ & $\Dastcmg\subset\Dcmg$, $\Dastten\subset\Dten$  \\
\bottomrule
\end{tabular}
\end{table}

To facilitate later experimental comparisons, we distilled four disjoint or nested subsets from the final 63,533-commit corpus ($\Dall$); their sizes and origins are summarised in \autoref{tab:rq_datasets}.

$\bullet$ RQ1 employs $\Dhuman$ (300 items sampled from $\Dcmg$) to quantify human–LLM agreement on our five binary metrics.

$\bullet$ RQ2 uses the full CMG-eval Dataset $\Dcmg$ (1,000 commits) for large-scale commit-message generation benchmarking.

$\bullet$ RQ3 relies on the Ten-category-eval Dataset $\Dten$ (1,160 commits) for ten-type classification evaluation.

$\bullet$ RQ4 performs two independent ablations: we randomly draw 200 commits from $\Dcmg$ to assess the impact of AST-level context on message generation, and another 200 commits from $\Dten$ to evaluate its effect on commit-type classification.

\section{Evaluation}
\label{sec:evaluation}
In this section, we report and analyze the experimental results to address the following research questions (RQs):

$\bullet$ \textbf{RQ1:} How consistent is the automated evaluation  of the binary evaluation criteria introduced in \S \ref{sec:evaluation_module} by LLM with human evaluation? 

$\bullet$ \textbf{RQ2:} How do existing tools and LLMs perform in the commit message generation task on ~\tool{}?

$\bullet$ \textbf{RQ3:} How do existing tools and LLMs perform in the commit classification task on ~\tool{}?

$\bullet$ \textbf{RQ4:} What is the impact of function-level contextual information on commit classification tasks and message generation tasks?

\subsection{RQ1: Consistency Between Automated LLM Evaluation and Human Evaluation of Binary Metrics}

As mentioned in §\ref{sec:evaluation_module}, we proposed a set of binary evaluation metrics to address the dependency of traditional evaluation metrics on reference texts and shift the evaluation mode from manual to automated assessment. Although this approach appears to save significant workload in evaluating related tools, we first need to validate the effectiveness of LLMs in this binary evaluation task before full-scale application.

We randomly draw 300 commits from the CMG experimental results of different tools/LLMs in §\ref{sec:CMG_performance} and evaluated them through a rigorous process involving three independent human reviewers (each with 3+ years of software development experience) and DeepSeek-V3. Each reviewer first scored every message on all five metrics according to a strict guideline; disagreements were then resolved in a moderated consensus session, yielding a single ground-truth label set. Inter-rater agreement among the three developers, quantified by pairwise Cohen’s Kappa , averaged 0.817, indicating “almost perfect” concordance. When the final human labels were compared with the fully automated scores produced by DeepSeek-V3, Cohen’s Kappa reached 0.849, surpassing the inter-human baseline. These results corroborate the high average Recall, Precision, F1, and Accuracy (\textgreater 0.9) reported in \autoref{tab:rq1_evaluation_consistency} and establish that the LLM-based assessment is not only consistent with, but slightly exceeds, human-level reliability.

\begin{table}[htbp]
\centering
\caption{Consistency results of binary metrics between LLM automated evaluation and human evaluation.}
\label{tab:rq1_evaluation_consistency}
\begin{tabular}{lccccc}
\toprule[1.2pt]
\textbf{Field} & \textbf{Recall} & \textbf{Precision} & \textbf{F1} & \textbf{Accuracy} \\
\midrule[1.2pt]
Rationality & 0.934 & 0.942 & 0.938 & 0.950 \\
Comprehensiveness & 0.931 & 0.938 & 0.934 & 0.943 \\
Non-redundancy & 0.960 & 0.845 & 0.899 & 0.857 \\
Authenticity & 0.924 & 0.959 & 0.941 & 0.903 \\
Logicality & 0.983 & 0.891 & 0.935 & 0.947 \\
\midrule
Average & 0.946 & 0.915 & 0.929 & 0.920 \\
\bottomrule[1.2pt]
\end{tabular}
\end{table}

\subsection{RQ2: Performance of Existing Tools and LLMs in CMG}
\label{sec:CMG_performance}

We selected four state-of-the-art (SOTA) CMG tools to cover major CMG paradigms: NNGen (retrieval-based)~\cite{liu2018neural}, FIRA (learning-based with AST)~\cite{dong2022fira}, CoRec (hybrid retrieval-generation)~\cite{wang2021context}, and KADEL (latest learning-based model achieving top performance)~\cite{tao2024kadel}.  Additionally, we include six popular LLMs from OpenRouter’s ranking (DeepSeek-V3, DeepSeek-R1, GPT-4.1, GPT-4o-mini, Gemini-2.0-flash, Claude-3.7-sonnet) to investigate their performance in CMG tasks. We used the CMG-eval Dataset from §\ref{sec:evaluation_module} for evaluation (focusing solely on the five reference-text-independent metrics).

For LLM evaluator, we used a zero-shot prompt (core excerpt below; full prompt available in our open-source repository):

\begin{tcolorbox}[colback=gray!10, colframe=gray!50, title={Core Prompt for LLM Evaluation}]
\scriptsize
You are a code reviewer...\\

You are given the following information:

- Modified files (with diffs)

- Possibly related issues, pull requests, or comments

- Possibly AST changes

- The generated commit message (CMG\_result)\\

You should assess the generated commit message from five perspectives:...(Rationality/Comprehensiveness/...)

\end{tcolorbox}

\begin{table*}[t!]
\centering
\caption{Performance comparison of existing tools, LLMs, and humans in CMG tasks.}
\label{tab:cmg_performance}
\small
\renewcommand{\arraystretch}{0.9}
\setlength{\tabcolsep}{3pt}
\begin{tabular}{lccccccccc}
\toprule[1.2pt]
\multirow{2}{*}{\textbf{Category}} & \multirow{2}{*}{\textbf{Name}} & \multicolumn{3}{c}{Traditional Evaluation Metrics} & \multicolumn{5}{c}{Binary Evaluation Metrics} \\
\cmidrule(r){3-5} \cmidrule(r){6-10} 
& & \textbf{BLEU} & \textbf{ROUGE-L} & \textbf{METEOR} & \textbf{Rationality} & \textbf{Comprehensiveness} & \textbf{Non-redundancy} & \textbf{Authenticity} & \textbf{Logicality} \\
\midrule[1.2pt]
\multirow{4}{*}{Existing Tools} 
& NNGen & 3.164 & 11.248 & 8.647 & \textbf{0.023} & 0.088 & 0.802 & \textbf0.238 & 0.198 \\
& CoRec & 4.142 & 14.887 & 10.562 & 0.020 & 0.084 & 0.810 & 0.345 & 0.231 \\
& FIRA & \textbf{5.374} & \textbf{19.421} & \textbf{12.359} & 0.013 & \textbf{0.278} & \textbf{0.914} & 0.693 & \textbf{0.629} \\
& KADEL & 0.216 & 6.207 & 1.511 & 0.017 & 0.094 & 0.544 & \textbf{0.761} & 0.017 \\
\midrule
\multirow{6}{*}{LLMs} 
& DeepSeek-V3 & 2.066 & 18.440 & 23.289 & 0.920 & \textbf{1.000} & 0.970 & 0.997 & \textbf{1.000} \\
& DeepSeek-R1 & 1.488 & 17.468 & 21.073 & 0.961 & \textbf{1.000} & 0.979 & 0.999 & \textbf{1.000} \\
& GPT-4.1 & 1.510 & 16.059 & 21.398 & \textbf{0.966} & \textbf{1.000} & \textbf{0.986} & \textbf{1.000} & \textbf{1.000} \\
& GPT-4o-mini & 1.848 & 16.169 & 21.571 & 0.959 & 0.997 & 0.971 & 0.985 & 0.998 \\
& Gemini-2.0-flash & \textbf{6.126} & \textbf{21.741} & \textbf{28.219} & 0.937 & 0.998 & 0.969 & 0.994 & 0.999 \\
& Claude-3.7-sonnet & 2.675 & 20.087 & 23.242 & 0.945 & \textbf{1.000} & 0.980 & 0.997 & \textbf{1.000} \\
\midrule
Humans & Human & - & - & - & 0.337 & 0.481 & 0.492 & 0.496 & 0.494 \\
\bottomrule[1.2pt]
\end{tabular}
\label{tab:CMG_performance}
\end{table*}

The experiment prioritized comparing tool and LLM performance on a new high-quality evaluation dataset. In CMG task, we did not optimize LLM prompts beyond instructing them to generate commit messages using provided diffs, issues, PRs, comments, and AST changes. The final results are presented in \autoref{tab:CMG_performance}.

Among existing tools, FIRA performed best, achieving the highest scores in six of eight metrics. Notably, it outperformed human developers in ``Non-redundancy'', ``Authenticity'', and ``Logicality'', suggesting high-quality training data and strong generalizability. KADEL, however, performed poorly, ranking lowest in five metrics. Its training data contained numerous default-format messages like  ``Merge pull request from...'' and  ``Update \{filename\}'', which describe commit actions rather than content~\cite{zhang2024automatic}, leading the tool to generate similarly meaningless outputs.

Across eight metrics, LLMs only scored lower than traditional tools in BLEU. Their performance on the five binary metrics was particularly notable: LLMs achieve \textgreater0.9 on each binary metric. Gemini-2.0-flash achieved the highest scores in BLEU, ROUGE-L, and METEOR, suggesting its outputs are most similar to high-quality human-written messages, while GPT-4.1 led in the remaining five metrics. This distinction arises because different models exhibit distinct characteristics in message generation: \textbf{traditional metrics} prioritize similarity to human-written references, while \textbf{binary metrics} emphasize comprehensiveness and completeness.

\begin{figure}[ht]
    \centering    \includegraphics[width=0.8\linewidth]{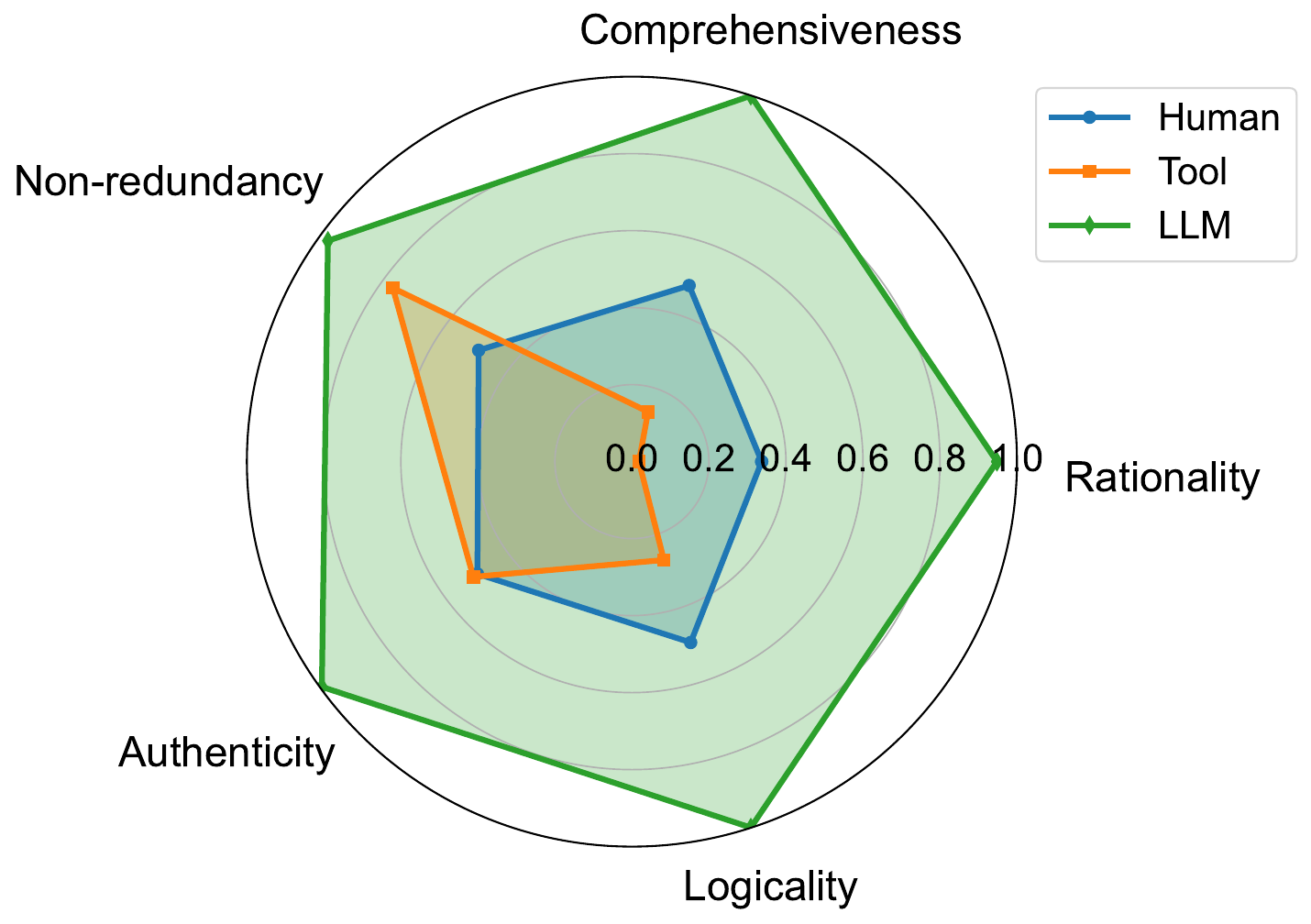}
    \caption{Compare the average scores of humans, LLMs, and tools across the five binary metrics.}
    \label{fig:human_tool}
\end{figure}

\autoref{fig:human_tool} compares the average scores of humans, LLMs, and tools across the five binary metrics. Notably, a random sample of 1,000 human-written messages from the full dataset performed modestly, partially explaining the tools’ subpar performance, low-quality training data limits tool capabilities, and their performance declines when evaluated on high-quality datasets. Zhang et al.’s experiments~\cite{zhang2024automatic} showed that retraining and testing tools on \textit{cleaned datasets} (with noisy commits removed) resulted in decreased evaluation scores, supporting this hypothesis. This indicates that conventional tools may have adapted to prevalent noise patterns in typical commit data.  Additionally, some evaluation data exceeded tools’ input length limits, resulting in truncated inputs that caused information loss and degraded performance.  

LLMs clearly hold an advantage in adapting to new scenarios, due to their support for extremely long contexts, which enables handling complex diffs and edge cases. However, ``Rationality'' and ``Non-redundancy'' remain areas for improvement, requiring future research on how to enable LLMs to accurately capture change rationales and refine message content. Meanwhile, BLEU, ROUGE-L, and METEOR remain relevant for LLMs, and efforts should focus on enhancing these metrics while maintaining performance on binary indicators to close the gap with human-written messages. 

This experiment demonstrates that \tool{}’s binary metrics sharply distinguish tool and LLM behaviors, while its rigorously curated CCS-compliant data lends renewed credibility to traditional scores, enabling a more holistic and trustworthy evaluation of commit message generation. 

\subsection{RQ3: Performance of Existing Tools and LLMs in CCS-Based Commit Classification}

Conventional Commits Specification (CCS) (§\ref{sec:cmg_background}) outperforms earlier taxonomies by mandating a machine-readable \verb|<type>| prefix that cleanly maps commits into ten fine-grained categories (\autoref{tab:commit-types}). Unlike ad-hoc project conventions, CCS is language-agnostic, documented in a single canonical spec, and natively supports downstream automation such as semantic versioning and changelog generation—features that make it the de-facto industry standard for commit classification.

We used the Ten-category-eval Dataset from §\ref{sec:evaluation_module} for evaluation. Zeng et al.~\cite{zeng2024first} are the first to study CCS-based commit classification, and their work remains the primary reference in this area. In addition to their classification tool (hereafter ``the classifier''), we included the six LLMs from  §\ref{sec:CMG_performance} to evaluate performance on the commit classification task. Results are shown in \autoref{tab:ten_category_performance}. 

\begin{table}[htbp]
\centering
\caption{Ten-category classification performance of Qunhong Zeng et al.'s method and LLMs.}
\label{tab:ten_category_performance}
\small
\footnotesize
\begin{tabular}{lcccc}
\toprule[1.2pt]
\textbf{Name} & \textbf{Precision} & \textbf{Recall} & \textbf{F1} & \textbf{Accuracy} \\
\midrule[1.2pt]
Qunhong Zeng et al & 0.580 & 0.473 & 0.457 & 0.473 \\
DeepSeek-V3 & 0.665 & 0.597 & 0.588 & 0.597 \\
DeepSeek-R1 & \textbf{0.675} & \textbf{0.638} & \textbf{0.621} & \textbf{0.638} \\
GPT-4.1 & 0.641 & 0.608 & 0.585 & 0.608 \\
GPT-4o-mini & 0.643 & 0.559 & 0.554 & 0.559 \\
Gemini-2.0-flash & 0.664 & 0.631 & 0.620 & 0.629 \\
Claude-3.7-sonnet & 0.668 & 0.609 & 0.606 & 0.609 \\
\bottomrule[1.2pt]
\end{tabular}
\end{table}

\begin{figure*}[t]
    \centering    \includegraphics[width=0.9\linewidth]{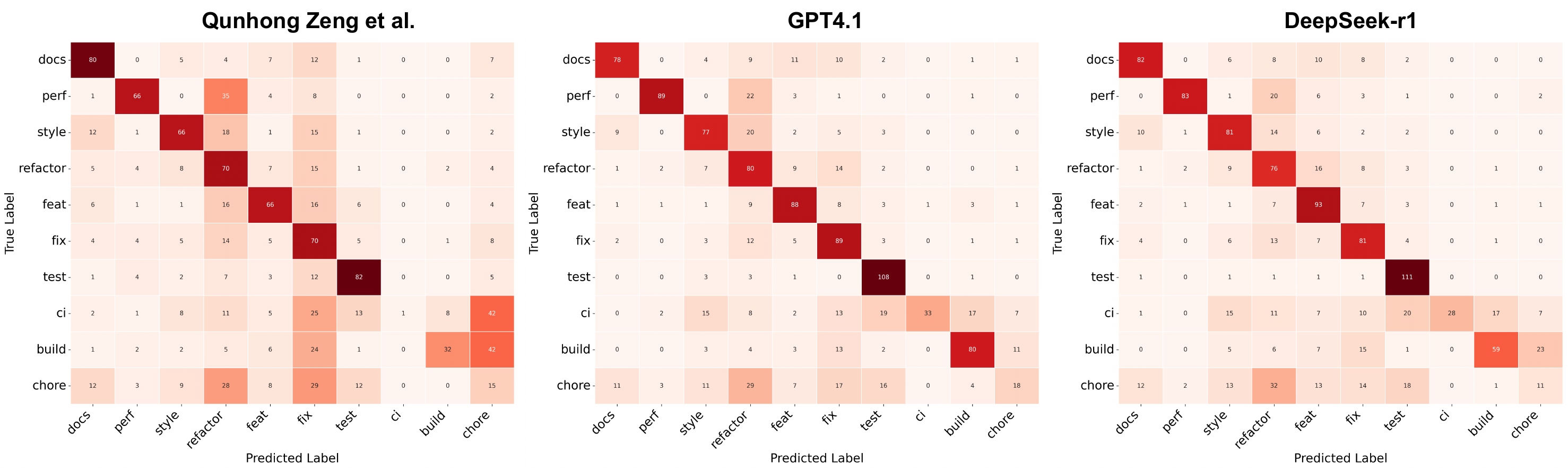}
    \caption{Compares the confusion matrices of the classifier, GPT-4.1, and DeepSeek-R1.}
    \label{fig:gpt_ds}
    \vspace{0.5em}
\end{figure*}

DeepSeek-R1 performed best among both the classifier and LLMs, achieving the highest scores in all four metrics with an accuracy of 0.6379. \autoref{fig:gpt_ds} compares the confusion matrices of the classifier and representative LLMs (GPT-4.1 and DeepSeek-R1), revealing frequent misclassifications among \verb|ci|, \verb|build|, and \verb|chore|. The \verb|build| category showed the most significant classification discrepancies. \verb|Ci| and \verb|build| involve system construction, testing, and deployment, often containing overlapping changes with other categories, leading to confusion. \verb|Chore| encompasses miscellaneous changes, but classifiers or LLMs unfamiliar with project structures struggle to determine what qualifies as ``miscellaneous'', compounded by subjective developer interpretations of the same changes. As \autoref{fig:type_distribution} previously showed, \verb|chore| commits comprise a substantial portion of real-world development, highlighting the need for clearer definitions, although standardization remains challenging. Beyond these three categories, \verb|perf|, \verb|style|, and \verb|fix| were frequently misclassified as \verb|refactor|. Further research is required for these categories prone to misjudgment or confusion. Overall, commit classification tools still have significant room for improvement. Addressing common misclassifications like those noted will be critical for future research. 

This experiment validate the unique contribution of \tool{}: by providing the first large-scale, CCS-compliant dataset with manually verified ten-category labels, we enable a controlled examination of classification boundaries that prior benchmarks could not support. The observed inter-category confusion underscores the need for precisely such high-quality data to drive future refinements in both rule-based and LLM-driven classifiers. 

\subsection{RQ4: Impact of Function-Level Contextual Information on Commit Tasks}

To investigate the impact of function-level contextual information on commit classification and message generation, we conducted two comparative experiments:

1) \textbf{First experiment (CMG task)}: We drew a random sample of 200 commits from the CMG-eval Dataset (\S \ref{sec:evaluation_module}). Using DeepSeek-V3, we generated messages for each commit with and without AST-level context. Human evaluators then pairwise compared the outputs along two dimensions—``what” (change content) and ``why” (change rationale). As shown in \autoref{tab:rq4_cmg_ast}, AST context yielded significantly better “what” descriptions (34 wins vs. 10 losses; Sign-test p = 0.0002). For “why”, AST still provided a slight advantage (9 wins vs. 2 losses), although the effect is weaker (p = 0.0327). Thus, AST-level information markedly improves the summarisation of \textit{what} changed, while its contribution to explaining \textit{why} the change was made is more modest. 

\begin{table}[htbp]
\centering
\caption{Comparison of ``what'' and ``why'' information generation with or without AST change in CMG task.}
\label{tab:rq4_cmg_ast}
\begin{tabular}{lcccc}
\toprule
\textbf{Metric} & \textbf{AST Better} & \textbf{No AST Better} & \textbf{Tie } & \textbf{Sign-test p-value}\\
\midrule
\textbf{What} & 34 & 10 & 156 & 0.0002\\
\textbf{Why} & 9 & 2 & 189 & 0.0327\\
\bottomrule
\end{tabular}
\end{table}

2) \textbf{Second experiment (commit classification task)}: We again sampled 200 commits from the Ten-category-eval Dataset (\S \ref{sec:evaluation_module}). Guided by the RQ3 results, we selected the top model (DeepSeek-R1), the worst model (DeepSeek-V3), and two median model (GPT-4o-mini, Gemini-2.0-flash). All four models were evaluated with and without AST information. \autoref{tab:rq4_classification_ast} shows AST data had negligible impact and slightly reduced LLM performance across metrics, suggesting it constitutes redundant information for classification tasks.

\begin{table*}[t!]
\centering
\caption{Ten-category classification performance of LLMs with and without AST change information.}
\label{tab:rq4_classification_ast}
\resizebox{0.65\linewidth}{!}{
\begin{tabular}{lcccccccc}
\toprule[1.2pt]
\multirow{2}{*}{\textbf{Name}} & \multicolumn{4}{c}{\textbf{Without AST}} & \multicolumn{4}{c}{\textbf{With AST}} \\
\cmidrule(r){2-5} \cmidrule(r){6-9} 
& Precision & Recall & F1 & Accuracy & Precision & Recall & F1 & Accuracy \\
\midrule[1.2pt]
DeepSeek-V3 & 0.6647 & 0.5974 & 0.5883 & 0.5974 & 0.6636 & 0.6078 & 0.6000 & 0.6078 \\
DeepSeek-R1 & 0.6752 & 0.6379 & 0.6214 & 0.6379 & 0.6600 & 0.6241 & 0.6018 & 0.6241 \\
GPT-4o-mini & 0.6433 & 0.5586 & 0.5536 & 0.5586 & 0.6429 & 0.5422 & 0.5335 & 0.5422 \\
Gemini-2.0-flash & 0.6635 & 0.6308 & 0.6203 & 0.6289 & 0.6654 & 0.6063 & 0.5980 & 0.6070 \\
\bottomrule[1.2pt]
\end{tabular}}
\end{table*}

Function-level contextual information aids ``what'' content generation in CMG tasks but is unnecessary for commit classification classification, where it may even degrade performance.

\section{Discusstion}
\label{sec:limitations}

\subsection{Case Study}

\begin{figure}[ht]
    \centering    \includegraphics[width=0.8\linewidth]{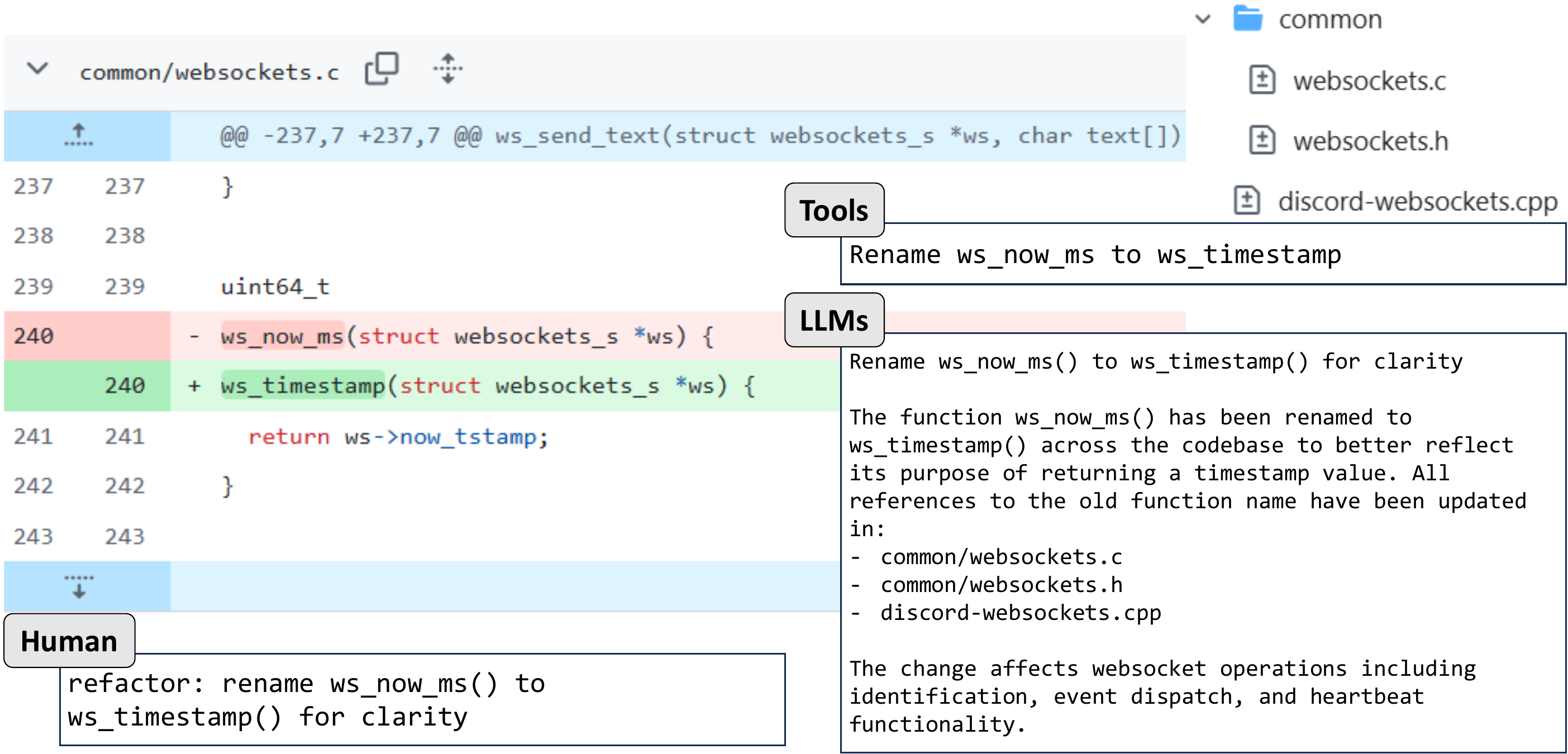}
    \caption{Compares the messages generated by humans, LLMs, and tools for the same commit.}
    \label{fig:huamen_tool_llm}
\end{figure}

Beyond the differences in metrics, there are other characteristics in the messages generated by humans, LLMs, and tools. \autoref{fig:huamen_tool_llm} compares the messages for the same commit: the human-generated message is precise and reasonable, the tool-generated message omits the change rationale (``for clarity''), while the LLM-generated message provides more comprehensive information, including all affected files and functions. Using current metrics, it is difficult to determine whether human-written messages or LLM-generated messages are better, which may depend on the specific needs of the repository. To address this, we may need to adjust LLMs according to real-world scenarios, requiring them to discard some ``less important details'' when necessary.

\subsection{Valid Analysis}

\textbf{(1) Internal Valid Analysis.}
Internal validity is threatened by the fact that approximately 67 \% of the collected commits omit explicit “why” information, thereby skewing model learning toward surface change descriptions; LLM-generated “what/why” annotations, though subject to 93.41 \% human verification, still propagate residual model bias; and the reliance on small, high-quality subsets for individual RQs limits generalizability to realistic, noise-laden repositories while revealing nonsignificant distinctions among LLMs under the proposed binary metrics, thus necessitating continued recourse to traditional similarity measures such as BLEU, ROUGE-L, and METEOR. Nevertheless, the ~\tool{} remains suitable for diverse commit-level tasks.

\textbf{(2) External Valid Analysis.}
External validity is constrained by evidence drawn exclusively from CCS-compliant repositories written in seven mainstream languages, leaving performance on non-CCS commits, niche languages such as Rust, and low-activity codebases unevaluated; moreover, while LLMs achieve high scores on the five binary metrics, their tendency to misclassify boiler-plate messages (e.g., “Merge pull request …”) as non-redundant suggests that current prompts may overfit to surface patterns rather than genuine content. Additionally, the reliance on the tree-sitter static analyser may introduce systematic deviations in the extraction of function-level information, further limiting the generalisability of our findings.

\section{Related Work}
\label{sec:related work}

\subsection{Benchmarks and CMG Tools}

CMG has evolved from rule-based~\cite{buse2010automatically} and statistical models~\cite{xu2019commit} to neural-based approaches~\cite{liu2018neural}, with growing focus on semantic richness. Among traditional tools, FIRA~\cite{dong2022fira} achieves strong performance by modeling fine-grained code edits with graph-based representations and incorporating a dual-copy mechanism. Such models often rely on AST-level information~\cite{wang2021context}, yet the integration of structural semantics remains tightly coupled with model-specific designs.
Regarding benchmarks, CommitBench~\cite{schall2024commitbench} provides a large-scale CMG dataset but lacks message quality annotations (e.g., ``what''/``why'') and standardized formatting checks, limiting its utility for fine-grained evaluation. Other datasets~\cite{mauczka2015dataset, eliseeva2023commit} support CMG research but are not aligned with practical labeling standards such as the CCS~\cite{conventionalcommitsConventionalCommits, zeng2024first}. Moreover, the absence of reliable CCS-labeled datasets and the semantic overlap between categories (e.g., \verb|fix| vs. \verb|ci|) pose persistent challenges for classification consistency.
These gaps highlight the need for a high-quality benchmark that not only supports CCS classification with reliable annotations but also enables fine-grained evaluation of commit messages in terms of structure and semantic adequacy.

\subsection{LLMs in CMG and Classification}

Recent studies have demonstrated that LMs significantly outperform traditional CMG models in terms of generalization and semantic reasoning~\cite{xue2024automated, wu2024commit, zhang2024using, imani2024context}. For example, Li et al.~\cite{li2024only} argue that code diffs alone are insufficient, and that incorporating structured reasoning processes can lead to more accurate message generation. Xue et al.~\cite{xue2024automated} further highlight that LLMs effectively capture both the ``what'' and ``why'' of a commit with minimal fine-tuning.
In parallel, researchers have begun applying LLMs to conventional commit classification. Zeng et al.~\cite{zeng2024first} introduced the first CCS classification benchmark, while Tong et al.~\cite{tong2025commit} enhanced classification performance using prompt tuning and external knowledge. Despite these advances, challenges such as ambiguous label boundaries and reliance on superficial cues persist.
To further improve LLM-based CMG and classification, recent work has explored integrating structural information such as ASTs~\cite{dong2022fira}, context-aware inputs~\cite{li2024only}, and prompting techniques including ICL, RAG, and MCP~\cite{zhang2024rag, xue2024automated}. However, systematically incorporating AST and history-aware representations into LLM workflows remains an open problem.
These findings collectively motivate the design of benchmarks and data protocols that align with LLM strengths, bridging structural code understanding with semantically rich commit reasoning.

\section{Conclusion}
\label{sec:conclusion}

In this paper, we present \tool{}, a comprehensive benchmark designed for both commit classification and commit message generation tasks. By integrating CCS (Conventional Commits Specification) and high-quality commit messages, we construct a large-scale, clean dataset covering multiple languages and repositories. We propose a new binary evaluation metric system that enables automated assessment of commit messages without relying on human-written references. Experiments demonstrate that LLMs outperform existing tools in both tasks, especially in message generation.

\bibliographystyle{ACM-Reference-Format}
\bibliography{main}

\end{document}